\documentclass[twocolumn,preprintnumbers,amsmath,amssymb]{revtex4}

\usepackage{graphicx}% Include figure files
\usepackage{dcolumn}% Align table columns on decimal point
\usepackage{bm}% bold math
\usepackage{color}
\usepackage[draft,english]{fixme}
\begin{document}

%\preprint{??????}

\title{Particle Interactions in Matter at the Terascale:  the Cosmic-Ray Experience}

\author{Spencer R. Klein} \affiliation{Lawrence Berkeley National Laboratory, Berkeley CA 94720 USA  \\ and the University of California, Berkeley, CA, 94720, USA}
 
%\date{\today}

\begin{abstract}

Cosmic-rays with energies up to $3\times10^{20}$ eV have been observed, as have as have astrophysical neutrinos with energies above 1 PeV.  In this talk, I will discuss some of the unique phenomena that occur when particles with TeV energies and above interact with matter.  The emphasis will be on lepton interactions.  The cross-sections for electron bremsstrahlung and photon pair conversion are suppressed at high energies, by the Landau-Pomeranchuk-Migdal (LPM) effect, lengthening electromagnetic showers.  At still higher energies (above $10^{20}$ eV), photonuclear and electronuclear interactions dominate, and showers become predominantly hadronic.  Muons  interact much less strongly, so can travel long distances through solids before losing energy.  Tau leptons behave similarly, although their short livetime limits how far they can travel.   The hadronic interaction cross-section is believed to continue to increase slowly with rising energy; measurements of cosmic-ray air showers support this prediction.

\end{abstract}

\maketitle

\section{Introduction}

Over the past 100 years, physicists have followed the cosmic-ray energy spectrum  upward in energy.  Modern large-acceptance detectors like the Flys Eye and Auger detector have observed cosmic-rays with energies up to $3\times10^{20}$ eV.   These cosmic-rays manifest themselves as particle showers, containing trillions of lower-energy particles, which cascade downward through the atmosphere.      These cascades develop through hadronic and electromagnetic interactions which convert particle energy into additional particles.    

Despite much effort, cosmic-rays are not yet well understood. We do not yet know their source(s) or their composition.  Are they protons or heavier nuclei?  The answer to this question requires a detailed understanding of the reactions that shape shower development.  Several experiments are trying to determine their origin, by searching for cosmic neutrinos produced in these accelerators \cite{Halzen:2008zz}.  This requires an accurate understanding of how leptons interact with matter.  

These neutrinos may be produced in astrophysical sources sources.  Many possible sources have been proposed, ranging from active galactic nuclei (galaxies with supermassive black holes at the center) to gamma-ray bursts (the collapse of supermassive stars, and/or collisions involving blacks holes and neutron stars).   One common, although not universal, feature of these models is that the neutrino spectrum is fairly hard, with the flux usually scaling as $E_\nu^{-2}$; this $E_\nu^{-2}$ is assumed in most experimental studies.  For an $E_\nu^{-2}$ spectrum, astrophysical neutrino detectors like IceCube and ANTARES are most sensitive to TeV and PeV neutrinos.  As will be discussed below, IceCube has observed two neutrino events with energies slightly above 1 PeV ($10^{15}$ eV), ushering in the era of terascale neutrino astrophysics.   

Neutrinos are also produced when cosmic-ray protons with energies above $4\times10^{19}$ eV interact with the  $3$ K microwave blackbody radiation and are photo-excited to the $\Delta^+$ resonance.  The  $\Delta^+$ decays, eventually producing neutrinos, mostly with energies in the $10^{17}$ - $10^{20}$~eV range.   Other experiments have searched for these  Greisen, Zatsepin and Kuzmin  (GZK) neutrinos, but are not yet sensitive enough to make definitive statements.  They have, however, set flux limits for neutrinos with energies up to $10^{25}$ eV.     
  
This writeup will review some of the unique physics that appears in particle interactions at the Terascale, and thus relevant to cosmic-ray and astrophysical neutrino studies.  I will emphasize phenomena that are unique to particle interactions in bulk matter, and are absent with individual atomic targets.  These phenomena are most prominent in reactions involving leptons, particularly electrons and photons.

\section{Electron and Photon Interactions}

At energies above $\approx 50$ MeV, electrons and photons interact predominantly through bremsstrahlung (‘braking radiation’) and pair production respectively.   In bremsstrahlung, an electron or positron interacts with a target nucleus and emits a photon, while in pair production, a photon interacts with a target and converts into an $e^+e^-$ pair.   These reactions share one key kinematic feature:  as the incident particle energy increases, the longitudinal momentum transfer ($q_{||}$) from the target nucleus decreases.  For bremsstrahlung of a photon
with energy $k$ from an electron with energy $E$ \cite{Klein:1998du},
\begin{equation}
q_{||} = \frac{km^2}{2E(E-k)}
\label{eq:lfzero}
\end{equation}
where $m$ is the electron mass. Per the uncertainty principle, the reaction is delocalized over a distance known as the formation length, 
\begin{equation}
l_f =\hbar/q_{||}  
\end{equation}
For pair production, the situation is similar.  For a pair with an invariant mass $M_{ee}$,
\begin{equation}
q_{||} = \frac{M_{ee}^2}{2k}
\end{equation}
with the formation length calculated the same way.  Most pair are produced with masses just above threshold, $M_{ee} \approx > 2m$ \cite{Klein:2004is}. 
Throughout this review, $E$ refers to electron energy, and $k$ refers to photon energy, for both bremsstrahlung and pair production.

For very high energy particles, the formation length can reach macroscopic distances.   For example, for a $10^{18}$~eV photon producing a 1 MeV (i.e. at threshold) pair, $l_f = 20$ cm.  For bremsstrahlung, even longer formation lengths are possible.  For a $10^{15}$ eV electron emitting a 1 GeV photon, the formation length is over a kilometer. When the formation length is longer than the inter-atomic distance a single electron or photon may interact simultaneously with multiple atomic targets, with the amplitudes adding, rather than the cross-sections.  These delocalized interactions can lead to rather non-intuitive behavior.

For bremsstrahlung, the photon radiation depends only on the total electron multiple Coulomb scattering (MCS)  in the formation length.  This mean scattering angle accumulates as the square root of the number of scatterers, so scales as $\sqrt{l_f}$.  Per standard electrodynamics \cite{Jackson}
\begin{equation}
\frac{d^2N}{dkd\Omega} = \frac{Z^2e^2\Gamma^4 |\Delta\vec{v}|^2 (1+\theta^4 k^4)}
{\pi^2 c^3 (1+\theta^2 k^2)^4}
\label{eq:radiation}
\end{equation}
where $N$ is the number of photons emitted,  $Z$ is the atomic number of the target atoms, $e$ the electrical charge, $\Gamma$ the projectile Lorentz boost, and $\Delta\vec{v}$ and $\theta$ the change in velocity and direction as the particle travels through the formation zone. When $\Gamma\theta < 1$, the $\sqrt{l_f}$ dependence leads to the same radiation as for independent scattering.  However, when $\Gamma\theta > 1$, the denominator in Eq. \ref{eq:radiation} rises rapidly, and the radiation is less than it would be if the electron interacted independently with each target atom \cite{LP}.  

\begin{figure}
\includegraphics[width=0.25\textwidth,angle=270]{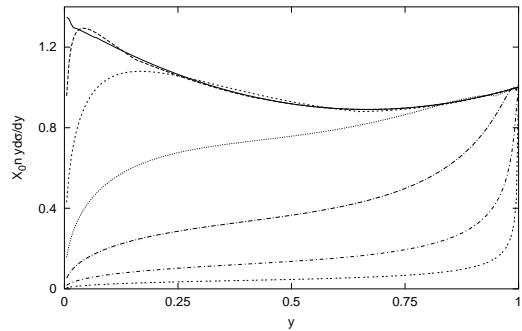}
\caption{The differential bremsstrahlung cross-section for different electron energies, in terms of $y=k/E$.  For lead targets, the curves correspond to electron energies of 10 GeV (solid line, top), 100 GeV (moving downward), 1 TeV, 10 TeV, 100 TeV, 1 PeV and 10 PeV (bottom curve).  The $y-$ axis shows the normalized cross-section times photon energy, $y d\sigma/dy \cdot 1/X_0$. From Ref. \cite{Klein:1998du}.}   
\label{fig:brem}
\end{figure}

One can calculate the suppression fairly simply using an expanded version of Eq. \ref{eq:lfzero} \cite{Klein:1998du}.  An additional term can account for the reduced longitudinal momentum due to multiple scattering (the momentum times $[1-\cos(\theta)]$).  Multiple scattering is usually treated with a Gaussian approximation.  Since the $\theta$ depends on $l_f$, and also partly determines it (by reducing the longitudinal momentum), some algebra leads to a quadratic equation for $l_f$.    Radiation is suppressed by the ratio of the formation length with multiple scattering to that without it.  When
\begin{equation}
\frac{k}{E} < \frac{E-k}{E_{LPM}}
\label{eq:ELPM}
\end{equation}
 bremsstrahlung is suppressed.  Here, $E_{LPM}=  7.7 X_0\ {\rm TeV/cm}$  is a material dependent constant; $X_0$ is the radiation length.   $E_{LPM}$ decreases rapidly with increasing density and atomic number, so LPM suppression is most important in dense media.  For example,  $E_{LPM}$ is 2.5 TeV in gold and 4.3 TeV in lead.  For water, important for astrophysical neutrino detectors, $E_{LPM} = 278$ TeV.  
When $k$ is smaller than in Eq. \ref{eq:ELPM}, the radiation is suppressed by $1/\sqrt{k}$, so $d\sigma/dk \approx 1/\sqrt{k}$.   
More detailed quantum-mechanical calculations have borne out the semi-classical derivation sketched out above.  Figure \ref{fig:brem} shows the bremsstrahlung differential cross-section for a variety of different electron energies.

\begin{figure}
\includegraphics[width=0.25\textwidth,angle=270]{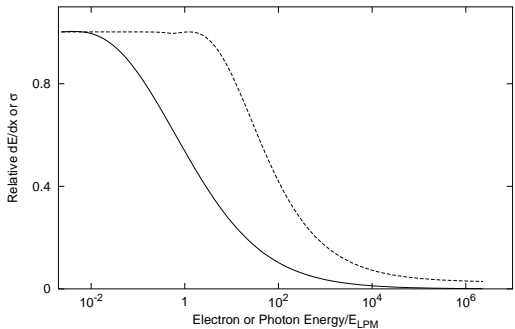}
\caption{The pair production cross-section (dashed line), and electron energy loss rate ($1/E dE/dx$, solid line), relative to the Bethe-Heitler cross-sections, as a function of photon or electron energy, divided by $E_{LPM}$.  From Ref. \cite{Klein:1998du}.}   
\label{fig:loss} 
\end{figure}

For $E \gg E_{LPM}$,  the cross-section is greatly reduced except when $k\approx E$,  and the electron typically transfers most of its energy to a photon.  Pair production behaves similarly.  For $k > E_{LPM}$, the cross-section is suppressed as $1/\sqrt{k}$.  Symmetric pairs are the most suppressed, leaving pairs where the electron or positron takes most of the energy.  Figure \ref{fig:loss} shows the pair production cross-section and normalized electron energy loss ($1/E\ dE/dx$) due to bremsstrahlung.   When $E \gg E_{LPM}$ an electron usually transfers most of its energy to a photon, which in turn transfers most of its energy to an electron or positron.  Since all of these interactions are subject to LPM suppression, the shower develops much more slowly than for unsuppressed (Bethe-Heitler) cross-sections. 

Bremsstrahlung photons can also interact with the medium, via forward Compton scattering.  In more classical language, the dielectric constant of the medium differs from one, slowing down the photon.  
One way to understand this effect is to treat the photon as having a mass, $\hbar\omega_p$, where $\omega_p$ is the plasma frequency of the medium.   In solid or liquid media, $\hbar\omega_p$ is 20-60 eV.   Creating a massive photon requires additional momentum transfer from the medium, reducing the formation length and suppressing the emission.  In solids, the photon mass significantly increases $q_{||}$ when $k/E < 10^{-4}$, and the bremsstrahlung cross-section is suppressed by the factor
\begin{equation}
\frac{k^2}{k^2+(\gamma\hbar\omega_p)^2}
\end{equation}
For $k < \gamma\hbar\omega_p$, the cross-section is strongly reduced, as $(\gamma\hbar\omega_p)^2/k^2$, and the bremsstrahlung cross-section scales as $d\sigma/dk \approx k$   The usual infrared divergence has naturally disappeared,  and the total bremsstrahlung cross-section is finite!  Of course, if the photon interacts in other ways, then these interactions can also suppress bremsstrahlung.  For optical or X-ray photons,  atomic physics plays a big role in regulating bremsstrahlung \cite{termik}.   At very high energies, photon pair production limits the formation length to the photon interaction length (including the LPM effect), suppressing bremsstrahlung by the ratio of the pair production length to the unsuppressed formation length.  An improved treatment adds an additional term to Eq.~\ref{eq:lfzero} to account for the finite pair production length \cite{GG}. 

In 1956, Migdal developed a fully quantum mechanical calculation of the suppression, using the Fokker-Planck equation to model the electron motion \cite{Migdal:1956tc}.   His calculation is commonly used today, although there are several newer and more sophisticated calculations, using a variety of different approaches \cite{Klein:2004is}.   Most of the newer calculations  can be applied to finite thickness targets; for targets thinner than the formation length, LPM suppression is reduced \cite{Klein:1998du,Kristoffer}.   For thin enough targets, the isolated-atom cross-section is recovered. They also include Coulomb corrections, which are significant for heavier nuclei, and a more accurate model of MCS. 

LPM suppression has been measured in a number of experiments involving cosmic-rays and accelerator beams.  The early (starting in the 1960's) cosmic-ray experiments suffered from poor statistics, and consequently limited precision, but one later cosmic-ray experiment observed an increase in length of electromagnetic showers in emulsion at energies around 100 TeV  \cite{Kasahara:1985ke}.

\begin{figure} [bt]
\includegraphics[width=0.5\textwidth]{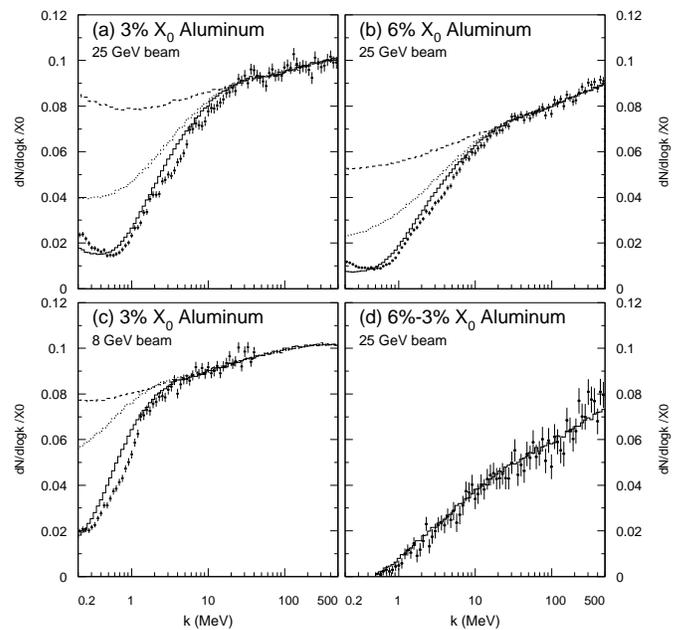}
\caption{The bremsstrahlung energy spectrum for 25 GeV and 8 GeV electrons traversing a 3\% $X_0$ and 6\% $X_0$ thick aluminum targets.   Photon energies from 200 keV to 500 MeV are shown, with 25 logarithmically spaced bins per decade of energy.  The dashed histogram is a simulation using the Bethe-Heiter cross-sections while the dotted line includes LPM suppression.  The solid histogram includes both LPM and dielectric suppression.   The upturn for photon energies below about 400 keV in both the data and simulation is mostly due to transition radiation. It is removed by subtraction (panel (d)).  From Ref. \cite{Anthony:1997ed}.}
\label{fig:E146}
\end{figure}

The first precision accelerator experiment was SLAC E-146, which sent 25-GeV and 8-GeV electron beams through targets of carbon, aluminum, iron, lead, tungsten and (depleted) uranium \cite{Anthony:1997ed}.  Figure \ref{fig:E146} shows an example of their data with 3\% and 6\% $X_0$ thick aluminum targets.  The $x$ axis shows the photon energy, from 200 keV to 500 MeV, with 25 logarithmically spaced bins per decade of energy.  Three theoretical curves are shown, but only the curve with LPM and dielectric suppression comes close to fitting the data.  The upturn below about 400 keV is due to transition radiation as the electrons enter and leave the target.

\begin{figure}
\center{\includegraphics[clip=,width=0.8\columnwidth]{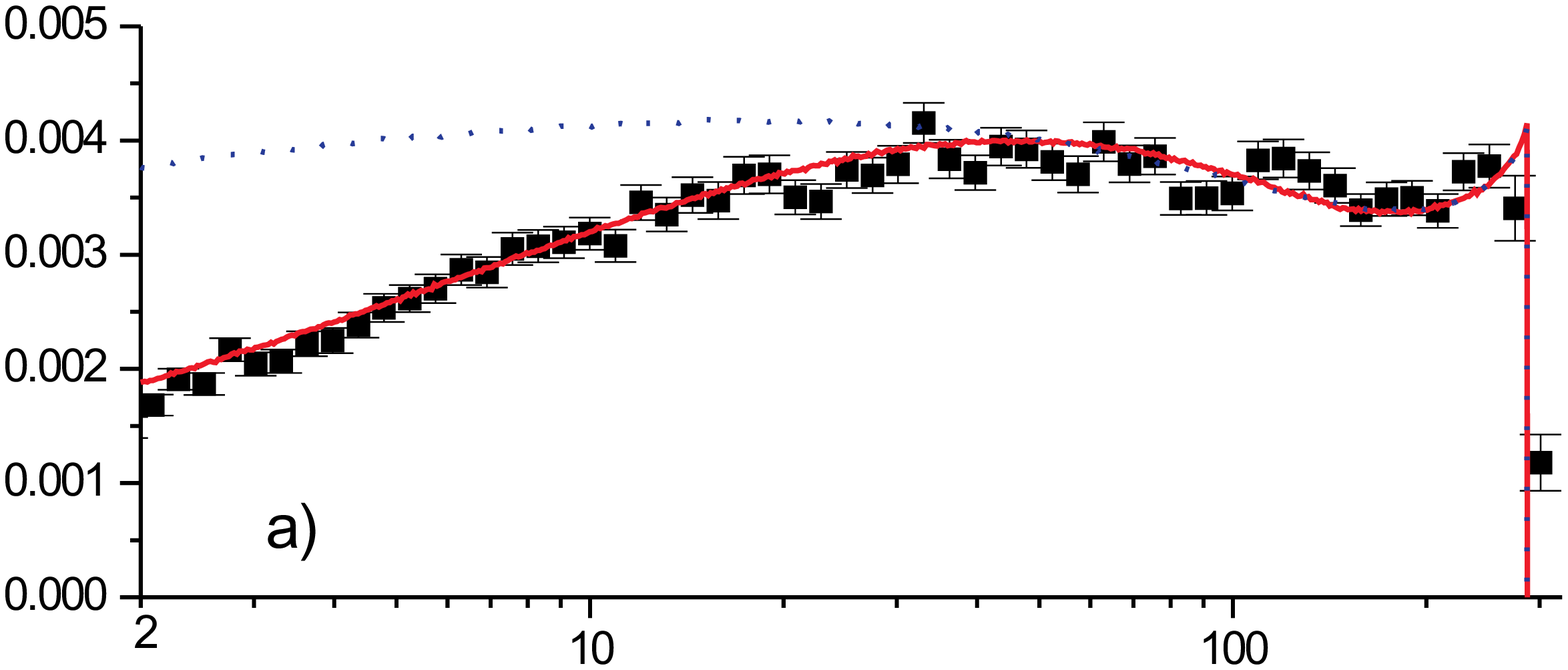}}
\center{\includegraphics[clip=,width=0.88\columnwidth]{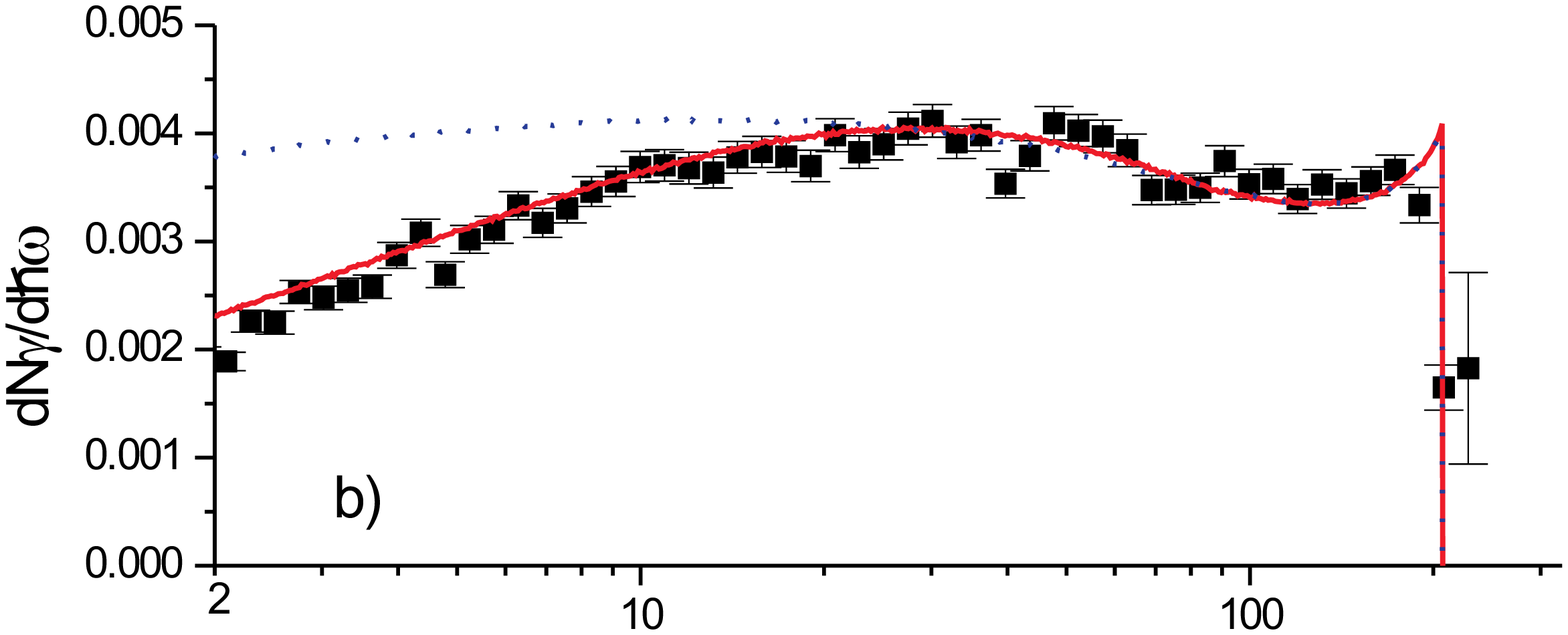}}
\center{\includegraphics[clip=,width=0.8\columnwidth]{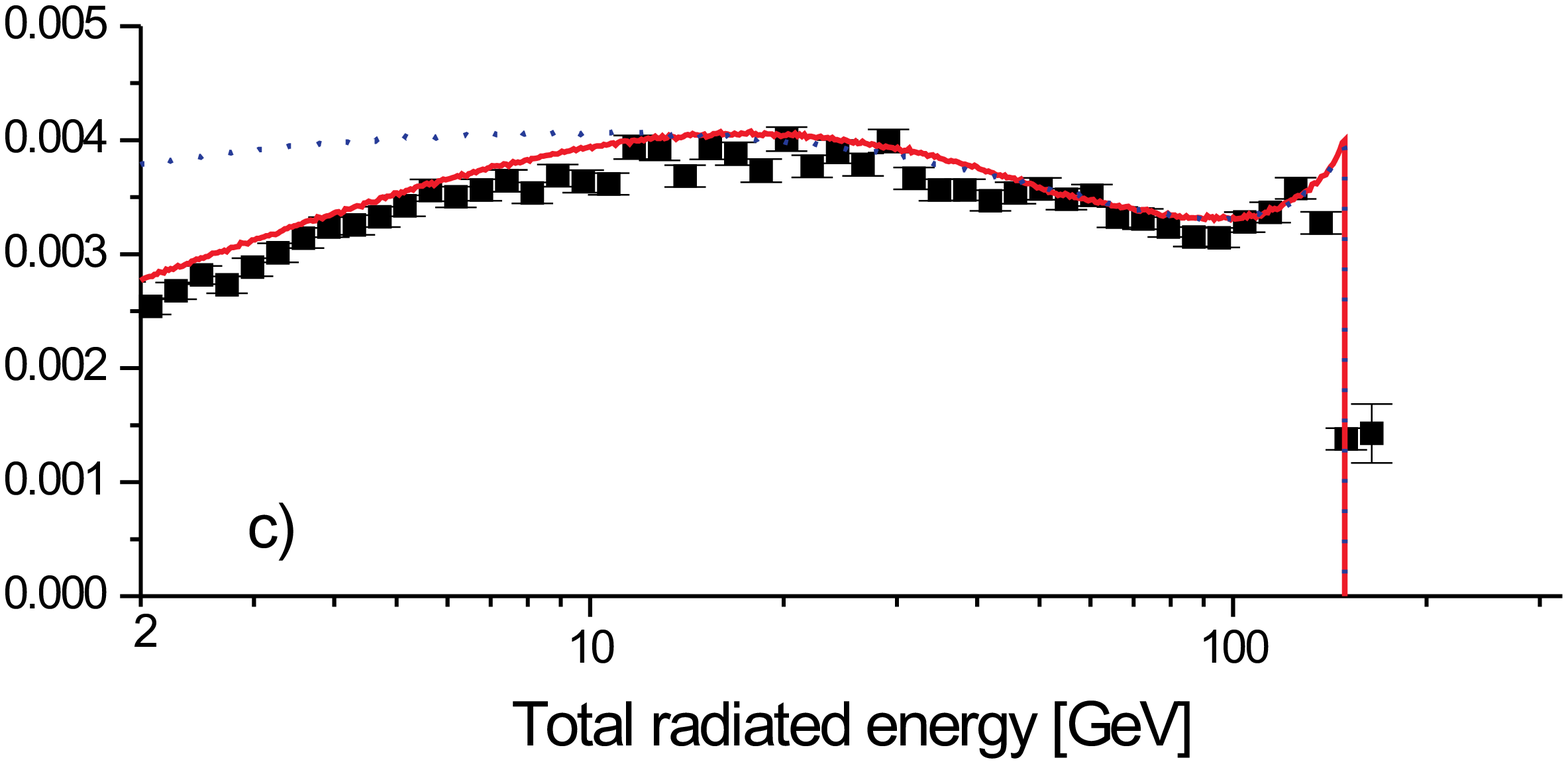}}
\caption{The bremsstrahlung energy spectrum for 287, 207 and 149 GeV electrons traversing a 0.435\% $X_0$ thick iridium target.   Photon energies above 2 GeV are detected.  For the 287 GeV electrons,  LPM suppression is visible up to $k/E\approx 0.12$.  From Ref. \cite{Hansen:2003bg}.}
\label{fig:CERN}
\end{figure}

More recently, a series of experiments at CERN has studied these electromagnetic phenomena, at higher beam energies, up to 287 GeV \cite{Hansen:2003bg}.  This is energetic enough that they were able to observe the  increase in radiation length as the electron energy approaches $E_{LPM}$.  Figure \ref{fig:CERN} shows their their data, from 287, 207 and 149 GeV electrons traversing an iridium target.    This group has also made some very interesting measurements of bremsstrahlung in crystalline targets, and also with thin targets \cite{Kristoffer}.  When the target was thinner than the formation length,  they observed the expected reduction in LPM suppression.

It should be mentioned that magnetic fields can also suppress bremsstrahlung.  Of course, the same fields will also induce synchrotron radiation, and a unified treatment is required.

\section{Photonuclear and Electronuclear Interactions}

At still higher energies photonuclear and electronuclear interactions become important.  In these reactions, photons (either real photons or virtual photons emitted by electrons) interact hadronically with target nuclei.  In the vector meson dominance picture, these photons fluctuate to quark-antiquark ($q\overline q$) pairs which then interact with the target. 
At very high energies,  the photon can also interact directly with a target quark. 
   
\begin{figure}
\includegraphics[width=0.5\textwidth]{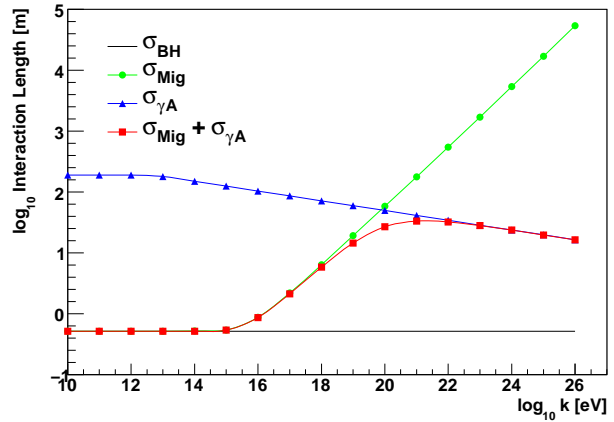}
\caption{Pair production and photonuclear interaction lengths for photons in an ice target.  Calculations are shown for the Bethe-Heitler cross-sections (no LPM suppression), LPM suppression of pair production, photonuclear interactions, and for all interactions.   There is  maximum in the interaction length around $10^{20}$ eV, where pair production becomes unimportant; at still higher energies, the photonuclear cross-section slowly rises.
From Ref. \cite{Gerhardt:2010bj}.}
\label{fig:photoncrosssections}
\end{figure}

When the LPM effect suppresses bremsstrahlung and pair production enough, photonuclear and electronuclear interactions dominate.   Figure \ref{fig:photoncrosssections} compares the pair production and photonuclear cross-sections in ice, with the cross-over around $10^{20}$ eV \cite{Gerhardt:2010bj}.   Different models predict photonuclear cross-sections that differ by roughly a factor of 4 \cite{Couderc:2009tq}, so the position of the cross-over is not well known.    Also, when  LPM suppression reduces the cross-section by more than the electromagnetic coupling constant, $\alpha\approx 1/137$, higher order processes, such as direct pair production, may become important; these require larger $q_{||}$, so are less subject to LPM suppression.  Unfortunately the higher order corrections to LPM suppression have not yet been calculated.

\begin{figure}
\includegraphics[width=0.5\textwidth]{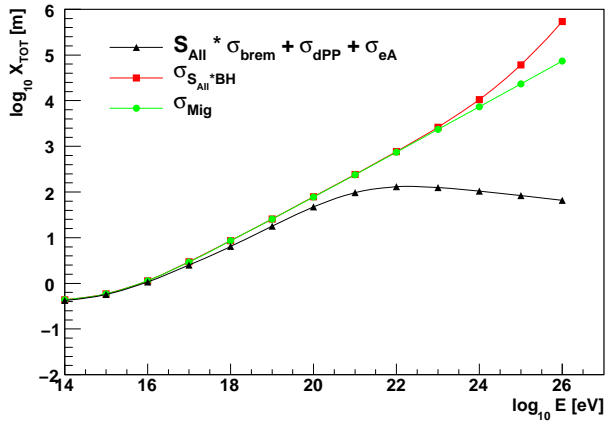}
\caption{The electron energy loss distance for electrons in an ice target, due to bremsstrahlung (with LPM and other suppression mechanisms) and electronuclear interactions.   There is  maximum in the distance around $10^{21}$ eV, where bremsstrahlung becomes unimportant; at still higher energies, electronuclear energy loss rises, and the distance drops. From Ref. \cite{Gerhardt:2010bj}.}
\label{fig:electronuclear}
\end{figure}

Electronuclear processes are similar.   Energy loss occurs when an electron emits a virtual photon which then interacts hadronically.  The loss rate is determined by the virtual photon energy spectrum and the hadronic cross-section.   Figure \ref{fig:electronuclear} shows the electronuclear energy loss distance (the distance for the electron to lose all but $1/e$ of its energy) due to bremsstrahlung (with LPM suppression) and electronuclear interactions.

\begin{figure}
\includegraphics[width=0.5\textwidth]{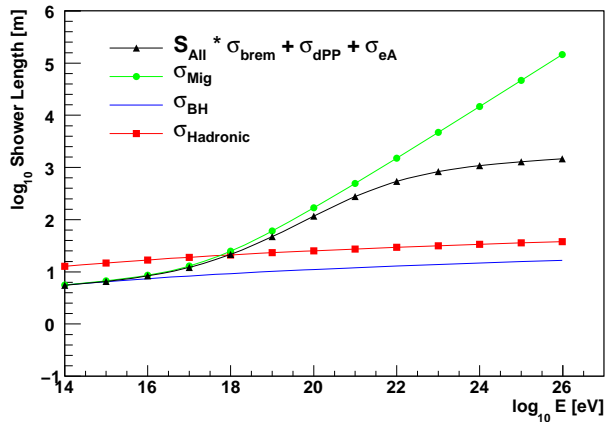}
\caption{The average total shower length for an electron-induced shower in an ice target, for Bethe-Heitler cross-sections, with LPM suppression, and including photonuclear interactions. Also shown, as a reference, is the shower length for a hadronic shower.  From Ref. \cite{Gerhardt:2010bj}.}
\label{fig:shower}
\end{figure}

Figure \ref{fig:shower} shows the average shower length for electron-neutrino ($\nu_e)$ induced showers calculated in a simple picture of shower development, with and without hadronic interactions \cite{Gerhardt:2010bj}.  The calculation includes direct radiation of electron-positron pairs, $e^-N\rightarrow e^-e^+e^-N$, which is normally unimportant for electrons, but can become significant when the LPM effect strongly suppresses bremsstrahlung.  Here, we neglect the roughly 20\% of its energy that a $\nu_e$ typically transfers to the struck hadron.
At energies below $E_{LPM}$, the shower length increases slowly, increasing roughly 1 $X_0$ for each doubling in shower energy.  However, when the LPM effect appears, the shower length increases more rapidly, as individual generations develop more slowly.  At energies around $10^{21}$ eV, hadronic interactions play a dominant role and the length increase moderates, around a value of 1 km, about 100 times larger than the length for a Bethe-Heitler shower.   In these regimes, the shower-to-shower variation in development is large, and one must account for these fluctuations in any study of energy spectra. 

At still higher energies, photonuclear interactions change character.  Photons which fluctuate to a $q\overline q$ pair can elastically scatter from the target, emerging as a real  vector meson, usually a $\rho^0$,  with the same quantum numbers as the photon.  When the formation length for a $q\overline q$ pair is larger than the typical internuclear separation, multiple nuclei can participate in the interaction.  Then, the $q\overline q$ scattering amplitudes add in-phase, and the cross-section rises.  Because $q\overline q$ pairs are much heavier than $e^+e^-$ pairs very high energies at required.  Still,  above $10^{23}$ eV, coherent conversion to $\rho^0$ mesons is predicted to be the dominant photonuclear interaction in solids \cite{Couderc:2009tq}.  At still higher energies, coherent converson continues, but heavier quarks play a significant role, leading to significant charm production \cite{Rogers:2009zza}.  This coherent conversion has many similarities with kaon regeneration.   

Heavier leptons are also of interest to cosmic-ray physicists.

\section{Muons and Taus}

Except for being heavier, and therefore unstable, muon and tau leptons behave similarly to the electron.   However, the mass does affect their interactions; for a given energy, they have a much smaller Lorentz boost than electrons.  They lose energy through the same processes as electrons - bremsstrahlung, direct pair production and electronuclear interactions - but the kinematics changes the relative importance of these processes.   Because of the reduced Lorentz boost,  their energy loss is much smaller than for electrons, and they can travel long distances through solids.  For example, a 100 TeV muon has an average range of about 10 km in ice.   Taus interact even less strongly, but their short livetime (0.29 ps) limits how far they can travel.   Because of the Lorentz boost, the LPM effect comes into play at much higher energies than for electrons.

Muon energy loss is of great interest to large neutrino detectors.  It is usually parameterized as
\begin{equation}
\frac{dE}{dx} = a + b E
\end{equation}
where $a \approx 240$ MeV/mwe accounts for the energy loss due to ionization, and $b\approx 0.347\times10^{-3}$/mwe  is the energy loss due to stochastic interactions: bremsstrahlung, direct pair production, and electronuclear interactions \cite{Beringer:1900zz,Chirkin:2004hz}.  Here, 'mwe' is the target thickness in equivalent meters of water; 1 meter of water corresponds to a column density of 100 g/cm$^2$.  This stochastic energy loss is proportional to the muon energy.   It dominates for muon energies above about 1 TeV, and the measured $dE/dx$ is often used to estimate the muon energy.     Because muons can lose significant energy in a single interaction, $dE/dx$ varies a lot, and a better estimate of the muon energy is achieved by dividing the muon path into independent segments, and only using the $\approx 60\%$ of the segments with the lowest $dE/dx$ in determining the energy \cite{IceCube:2012re}. The mathematical picture is similar to particle energy loss in wire chambers.  

High-energy neutrino interactions are studied by a number of detectors.  The largest is the 1 km$^3$ IceCube neutrino observatory, located in the South Pole icecap \cite{Halzen:2010yj}.  Its 5,160 buried optical sensors observe Cherenkov radiation from the charged particles produced when high-energy (above 10-100 GeV) neutrinos interact in the Antarctic ice.   The different lepton interactions and limited tau lifetime lead to very different topologies for electron, muon and tau neutrinos. 

\begin{figure}
\center{\includegraphics[clip=,width=0.6\columnwidth]{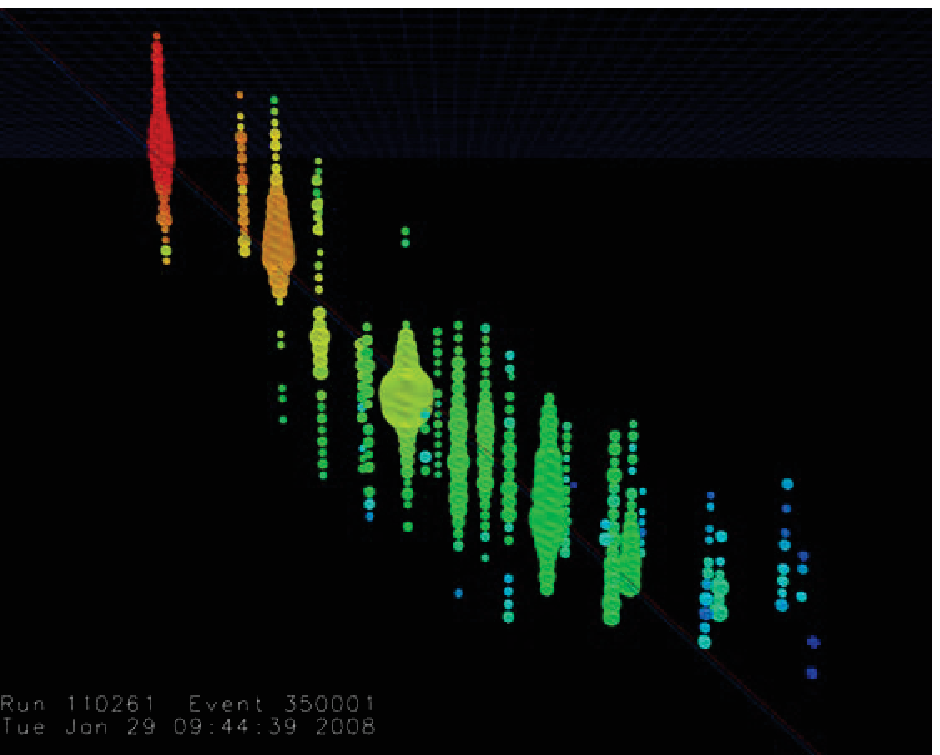}}
\center{\includegraphics[clip=,width=0.6\columnwidth]{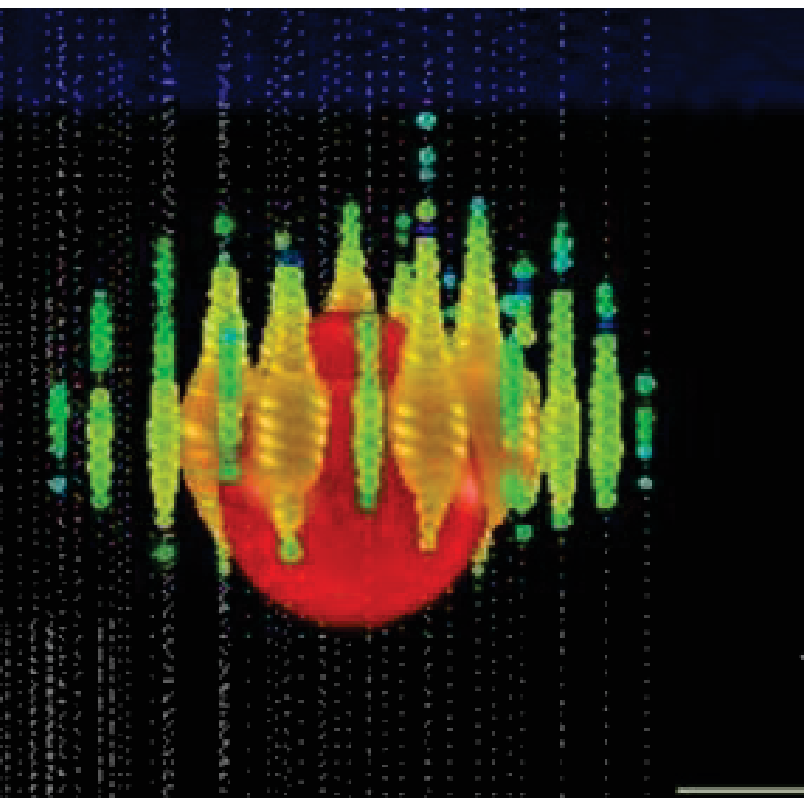}}
\center{\includegraphics[clip=,width=0.6\columnwidth]{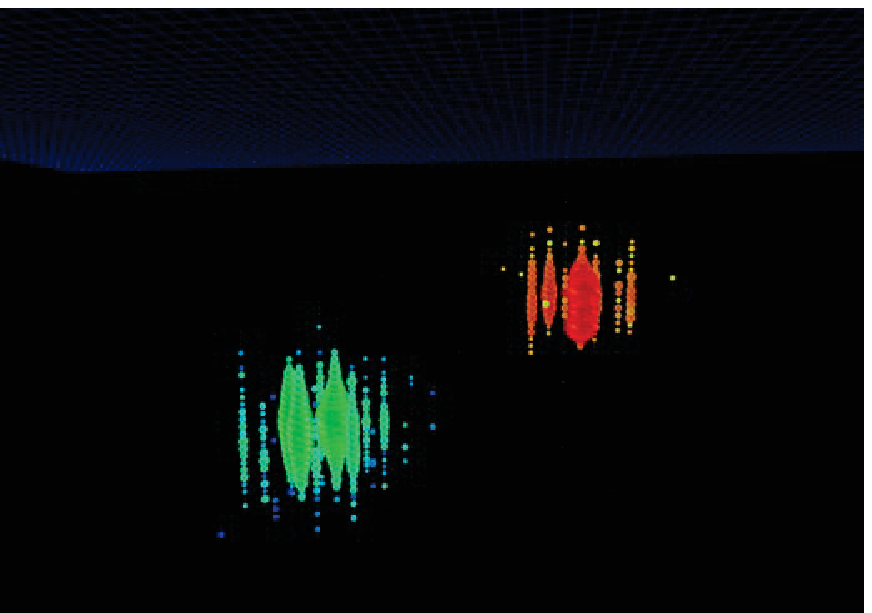}}
\caption{Three different types of neutrino events as observed in the 1 km$^3$ IceCube neutrino detector.   Each dot shows 1 of the 5,160 buried optical sensors mounted on 86 vertical strings, with a 17 m spacing between sensors.  The circles show the hits sensors, with the colors showing the arrival times of the first photon, from red (earliest) to blue (latest).  The size of the spheres indicates the number of photons that that sensor observed.  From top to bottom, there is a muon (or muon bundle) \cite{Halzen:2010yj}, a cascade \cite{Aya}, and a simulated tau double-bang event \cite{Halzen:2010yj}.  The cascade has an energy of about 1 PeV.}
\label{fig:IceCube}
\end{figure}

 Figure \ref{fig:IceCube} shows the three different topologies.  The top panel shows a muon (or multiple muon bundle).  It is observed for more than 1 km, allowing for a good $dE/dx$ measurement.  Over 1 km, a 10 TeV muon loses an average of 38\% of its energy traversing the 1 km of ice (the density of ice is 92\% of that of water).  The middle panel shows a shower from a neutrino interaction; this could either be an electromagnetic shower from a $\nu_e$ charged current interaction, or a neutral current interaction of any neutrino flavor.  Since the electromagnetic and hadronic shower lengths are short (39 cm and 92 cm respectively), and the energy is low enough that the LPM effect does not significantly affect shower development.  The shower deposits most of its energy in a small volume, per Fig. \ref{fig:shower}, but the Cherenkov light is visible from several hundred meters away.  The estimated $\nu$ energy is over 1 PeV \cite{Aya}.  The bottom panel shows a simulated $\nu_\tau$ double bang event \cite{Learned:1994wg}, with a hadronic shower produced when the $\nu_\tau$ interacts with the ice, transferring energy to the struck nucleon, a largely invisible $\tau$ track, and a second cascade where the tau decays.   

\section{Hadronic Interactions}

High-energy hadronic interactions are much less well understood than electromagnetic reactions.  Perturbative quantum chromodynamics (pQCD) predicts the behavior of hard interactions (those involving a large momentum transfer), but the coupling constant is large at low momentum transfers, so perturbative calculations do not give reliable predictions.  Unfortunately, most interactions are at low momentum transfers.  In this region, different calculations (and the resulting Monte Carlo simulation codes) rely on phenomenological models involving Pomeron exchange.  These simulations are tuned to generally agree with the available accelerator data, but diverge significantly at higher energies.  The one generally reliable prediction is that cross-sections should rise relatively slowly with increasing energy \cite{Block:2006hy}.    

Experimentally, the Auger collaboration has used cosmic-ray air showers to measure the proton-air cross-section at an average energy of $10^{18.24}$ eV, corresponding to a proton-proton center of mass energy of 57 TeV, far above the energy of the Large Hadron Collider (LHC).  They did this using a wide-field optical detector which imaged the fluorescence of atmospheric nitrogen induced as the shower propagated downward, measuring the altitude $X_{\rm max}$ at which the shower had the highest particle density.  Iron and other heavy (compared to protons) nuclei are larger, and tend to interact higher in the atmosphere.  Since their energy is split among many nucleons, the showers develop faster.  So, the collaboration selected a sample of showers that developed deep in the atmosphere, and used that to measure $\sigma_{\rm p-air} = 505 \pm 22 {\rm (stat.)}^{+22}_{-36} {\rm (syst.)}$ mb \cite{Collaboration:2012wt}.   This is about twice the cross-section measured at lower energies.  The collaboration also used a Glauber model to infer the proton-proton cross-section, and measured a value about 30\% higher than was seen at the LHC, in agreement with the scaling predicted by models where the cross-section grows relatively slowly with energy.

In a related analysis, the Auger collaboration studied the distribution of $X_{\rm max}$ as a function of shower energy.  They observed a change in slope in the elongation rate, $dX_{\rm max}/dE_{cr}$ at an energy of $10^{18.4}$ eV \cite{PierreAuger:2011aa}.  When compared with air shower simulations using two different hadronic interaction models, the kink is compatible with a change in cosmic-ray composition from mostly protons to mostly iron, over the energy range from $10^{18.4}$ to $10^{19.4}$ eV.  In the same energy range, the width of the $X_{\rm max}$ distribution narrows dramatically, with the root-mean-square of the distribution narrowing by about a factor of two; this is again compatible with a dramatic change in composition.  Although these changes are not observed by the Hires Flys Eye experiment \cite{Abbasi:2009nf}, that experiment has somewhat lower statistics.   If the changes in elongation rate and RMS variation in $X_{\rm max}$ are real, one cannot exclude the possibility of new hadronic physics, such as gluon recombination involving gluons at low Bjorken$-x$ values, or (in other words), parton saturation \cite{Pajares:2000sn}. 

\section{Coherent Cherenkov radiation}

Although it is not a significant contributor to particle energy loss, coherent Cherenkov radiation has become an  important tool for detecting ultra-high energy neutrinos. It occurs because electromagnetic and hadronic showers include more electrons than positrons.  The electron excess is due to two processes: a positron in the shower can annihilate on an electron in the target, removing its charge, or a photon in the shower can Compton scatter a target electron into the shower.  Both of these processes primarily affect low-energy particles, which are the majority of the particles in the shower; the net effect is about a 20\% excess of electrons over positrons \cite{Askaryan}.  When viewed at a wavelength that is larger than the transverse size of the shower (more precisely, at the Cherenkov angle), individual shower particles cannot be observed; instead the fields depend only on the net charge in the shower and the emission is coherent.  This condition is satisifed for radio waves, with frequencies up to about 1 GHz in ice, and several GHz in rock.   Below these frequencies, the electric field amplitude scales as the square of the neutrino energy.  At neutrino energies above about $10^{17}$ eV, this leads to a large signal.  Near the maximum frequency, most of the radio waves are emitted near the Cherenkov angle (forming a cone), but, at lower frequencies, the emission becomes more isotropic. 

Radio emission has also been observed from cosmic-ray air showers.  These showers have a much larger lateral extent, so the useful frequencies are much lower \cite{Falcke:2008qk}.  In addition to coherent Cherenkov radiation, there is also synchrotron radiation from low-energy electrons spiraling in the Earth's magnetic field, plus a dipole field as positively and negatively charged particles separate in the Earth's magnetic field; these latter two components are larger than the coherent Cherenkov radiation and vary, depending on the angle between the shower direction and the Earth's magnetic field \cite{Ardouin:2009zp}. 

Radio Cherenkov is the only technique that has been demonstrated to scale to the very large detector volume required for a next-generation neutrino detector.  Many existing experiments have already searched for radio emission from neutrino induced showers \cite{Klein:2010wf}.   Antarctic ice and the lunar regolith are common choices.  The ANITA balloon-borne detector has twice circled Antarctica at an altitude of around 35,000 m, with 32 (or 40) horn antennas searching for radio emission out to the horizon, up to 650 km away.   ANITA has not yet seen a signal, but it has set significant limits on the possible flux of neutrinos with energies above about $10^{19}$ eV \cite{Gorham:2010kv}.    

Looking ahead, two groups are prototyping arrays of antennas which will be co-located in the active volume, giving them much lower neutrino energy thresholds, of order $10^{17}$ eV.  The ARA collaboration proposes to deploy an array of antennas in shallow holes at the South Pole \cite{Allison:2011wk}.  The ARIANNA collaboration plans to deploy an array of antennas just below the surface, on the Ross Ice Shelf \cite{Klein:2012bu}.  Although the ice there is warmer, and so more absorptive, the interface between the ice shelf and the Ross Sea below it acts as a mirror for radio waves, reflecting Cherenkov radiation from downward-going neutrinos to the surface.  Both experiments are easily scalable, and both groups plan to instrument a volume larger than 100 km$^3$, enough to test models of cosmic neutrino production. 

\section{Conclusions}

Cosmic-ray physics tests our understanding of particle interactions at energies up to $10^{20}$ eV.  A these very high energies, new electromagnetic phenomena appear, in which leptons and photon interactions with bulk matter can be very different from their interactions with isolated atoms.   The LPM effect and other phenomena suppress bremsstrahlung and pair production, and, at high enough energies, electron and photon interactions become predominantly hadronic.    The character of leptonic interactions depends heavily on the lepton mass and lifetime, and we use these differing characters to determine the flavor of neutrinos that interact in large neutrino observatories.   Hadronic interactions are not well understood, but we have used cosmic-rays to measure interaction cross-sections at energies above those accessible at the large hadron collider.   Fortunately, some of the new interactions that come into play at very high energies create useful signatures for observing the particles that produce them; coherent radio Cherenkov emisison is the technology of choice for the next generation of neutrino detectors that will use instrumented volumes above 100 km$^3$ to search for neutrinos with energies above $10^{17}$ eV.

\section*{Acknowledgements}

Lisa Gerhardt provided useful comments on the manuscript.   This work was funded in part by the U.S. National Science Foundation under grant 0653266 and the U.S. Department of Energy under contract number DE-AC-76SF00098.

\end{document}